\documentclass[
reprint,
 amsmath,amssymb,
 aps,
prd,
]{revtex4-2}

\usepackage{graphicx}
\usepackage{bm}
\usepackage{hyperref}

\begin{document}


\title{Confusion noise from Galactic binaries for Taiji}

\author{Chang Liu$^{1,2}$}
\email{liuchang@ucas.ac.cn}

\author{Wen-Hong Ruan$^{1,2}$}
\email{ruanwenhong@ucas.ac.cn}

\author{Zong-Kuan Guo$^{3,2,1}$}
\email{guozk@itp.ac.cn}

\affiliation{$^1$School of Fundamental Physics and Mathematical Sciences, Hangzhou Institute for Advanced Study, University of Chinese Academy of Sciences, Hangzhou 310024, China}
\affiliation{$^2$School of Physical Sciences, University of Chinese Academy of Sciences, No.19A Yuquan Road, Beijing 100049, China}
\affiliation{$^3$CAS Key Laboratory of Theoretical Physics, Institute of Theoretical Physics, Chinese Academy of Sciences, P.O. Box 2735, Beijing 100190, China}

\begin{abstract}
Gravitational waves (GWs) from tens of millions of compact binaries in our Milky Way enter the milli-Hertz band of space-based detection. 
The majority of them cannot be resolved individually, resulting in a foreground confusion noise  for Laser Interferometer Space Antenna (LISA).
The concept of Taiji mission is similar to LISA's with slightly better sensitivity, which means that the galactic GW signals will also affect the detection with Taiji.
Here we generate the GW signals from 29.8 million galactic binaries for Taiji and subtract the `resolvable' sources.
The confusion noise is estimated and fitted in an analytic form with 6-month, 1-year, 2-year and 4-year observation time.
We find that the full sensitivity curve is slightly lower for Taiji than for LISA at frequencies of $\leq 0.8$ mHz and around 2~mHz. 
For a 4-year lifetime, more than 29 thousand sources are resolvable with Taiji.
Compared to LISA, Taiji can subtract $\sim 20 \%$ more sources and the distribution of them in our Milky Way is consistent with that of the resolvable sources with LISA.
At frequencies around 2~mHz or with the chirp masses ranging from  $0.2 M_\odot$ to $0.4 M_\odot$, more sources become resolvable with Taiji.
\end{abstract}

\maketitle


\section{\label{sec:intro}Introduction}

Space-based GW detectors, such as LISA, Taiji and TianQin, will open the milli-Hertz window for GW astronomy in 2030s~\cite{LISA:2017pwj, Hu:2017mde, TianQin:2015yph}. 
Sources with a wide range of masses and mass ratios may enter the frequency band from 0.1~mHz to 0.1~Hz, which include massive black hole binaries, compact binaries in the Milky Way, extreme mass ratio inspirals (EMRIs), stellar-origin black hole binaries, etc~\cite{Klein:2015hvg, Hils:1990vc, Nelemans:2001nr, Marsh:2011yj, Amaro-Seoane:2007osp, Babak:2017tow, Gerosa:2019dbe, Moore:2019pke, Toubiana:2020drf, Gair:2022knq}.

Another crucial goal for space-based detection is stochastic GW backgrounds of different origins.
In the milli-Hertz band, GW backgrounds may come from the early Universe such as cosmological phase transitions, inflationary reheating, the interactions of cosmic string, etc~\cite{Caprini:2015zlo, Christensen:2018iqi, Khlebnikov:1997di, Easther:2006gt, Auclair:2019wcv, Gair:2022knq}.
From the signals observed by LIGO and Virgo, the extragalactic binary black holes and binary neutron stars also contribute a power-law stochastic GW background for space-based detection~\cite{LIGOScientific:2016fpe, Chen:2018rzo}.
Furthermore, since the detection rate of EMRIs is very uncertain~\cite{Babak:2017tow}, under the most optimistic astrophysical assumptions, the population of the unresolved EMRIs could produce a stochastic background exceeding the instrument noise of LISA~\cite{Bonetti:2020jku}.

Based on the astrophysical population models~\cite{Korol:2020lpq, Korol:2021pun}, tens of millions of compact binaries in our Milky Way, so-called \textit{galactic binaries} (GBs), may simultaneously emit GWs in the frequency band from 0.1~mHz to 10~mHz~\cite{Evans:1987qa, Nissanke:2012eh, Cornish:2017vip, Karnesis:2021tsh}.
Only a small number of them, about ten or twenty thousand, are resolvable with LISA~\cite{Nissanke:2012eh, Cornish:2017vip, Karnesis:2021tsh}.
The majority of these sources are unresolved and form a stochastic `galactic foreground' or `confusion noise' for LISA~\cite{Nissanke:2012eh, Crowder:2004ca, Cornish:2017vip, Karnesis:2021tsh}.
The detection of the signals from our galaxy may provide information about the evolution and distribution of compact binaries in the Milky Way, which is one of the main targets of space-based detection~\cite{LISA:2017pwj, Hu:2017mde, Adams:2012qw, Gair:2022knq, Georgousi:2022uyt}.

The simulations and analyses for LISA have shown that the confusion noise is dominant over the instrument noise around 1~mHz~\cite{Nissanke:2012eh, Cornish:2017vip, Karnesis:2021tsh}.
Since the detection frequency band of TianQin is slightly higher than that of LISA and Taiji, the effect of the galactic confusion noise can be largely ignored for TianQin~\cite{Huang:2020rjf}. 
The triangle configuration of Taiji is similar to LISA's but with a longer arm-length and different heliocentric orbit~\cite{Hu:2017mde, LUO2020102918}. 
The centroid of the constellation leads the Earth by twenty degrees but for LISA the centroid trails the Earth by twenty degrees~\cite{LISA:2017pwj, LUO2020102918}.
In addition, the noise level of Taiji is slightly lower than LISA's~\cite{LISA:2017pwj, LUO2020102918}.
The differences in the concept of the two detectors indicate that the confusion noise from GBs for Taiji should be investigated separately.

In this paper, we use a catalog of GBs~\cite{Korol:2020lpq} to generate the signals from our galaxy for Taiji and subtract the resolvable sources from it.
Then, the confusion noise for Taiji is fitted in an analytic form for different observation time in Section~\ref{sec:sctaiji}.
In Section~\ref{sec:resolvable} we compare the number and the distribution of the frequencies, chirp masses and sky locations of the resolvable sources for Taiji with that of the sources for LISA.
Finally, we summarise our results in Section~\ref{sec:summary}.
Here we use units with $c=1$, where $c$ is the speed of light.

\section{\label{sec:sctaiji}Confusion noise for Taiji}

We follow the procedure widely used for LISA~\cite{Timpano:2005gm, Cornish:2017vip, Karnesis:2021tsh} to simulate the foreground signals and estimate the confusion noise for Taiji.
For the population of GBs, we use the catalog of the LISA Data Challenge (LDC) Radler dataset which contains about 29.8 million GB sources in mHz band~\cite{Baghi:2022ucj, Korol:2020lpq}.

\begin{figure}
  \includegraphics[width=0.9\linewidth]{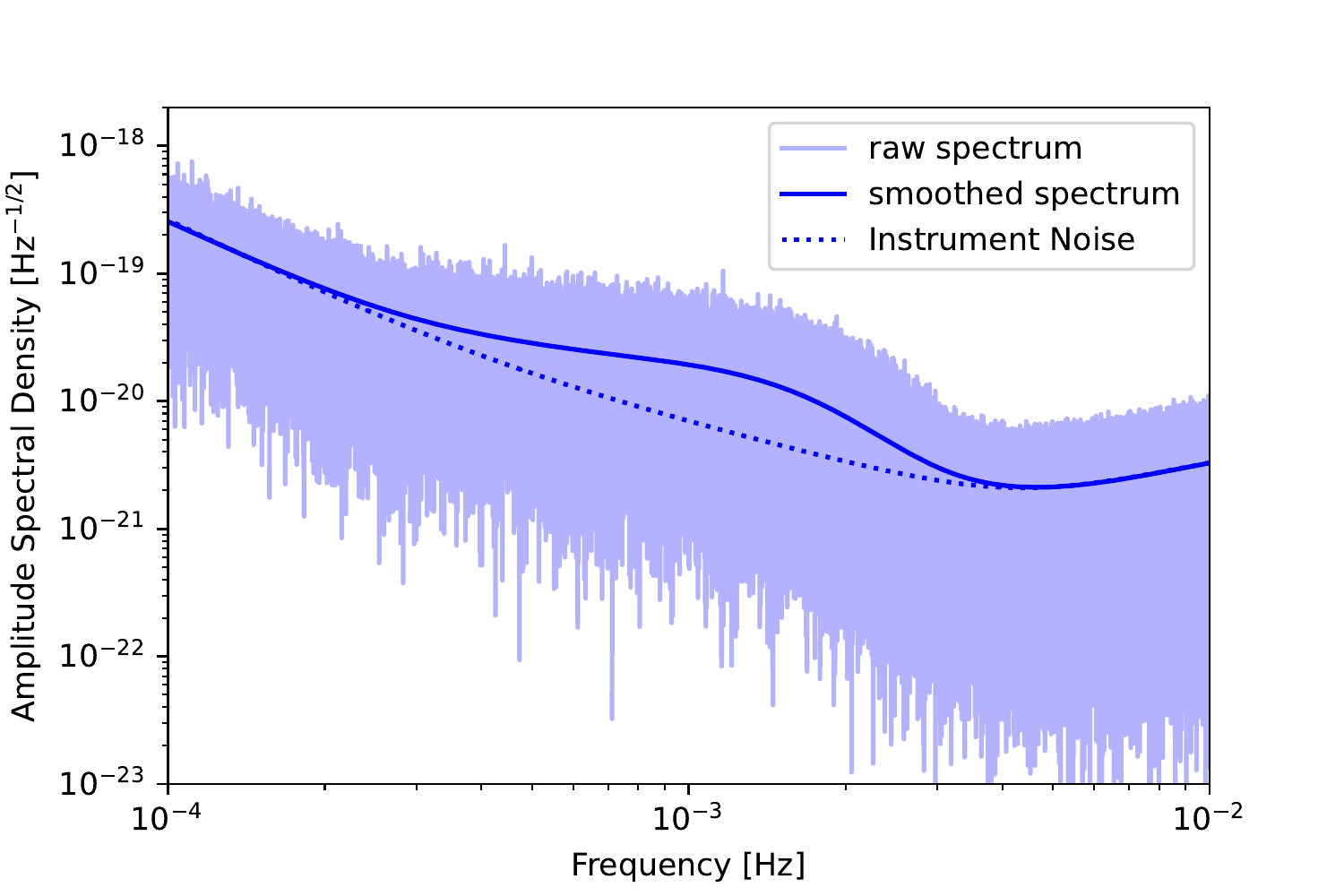}
  \caption{\label{fig:pnpcX} The amplitude spectral density in the X channel of Taiji with a 4-year mission duration. The blue curve shows the estimated smoothed spectrum obtained from the BayesLine algorithm. The instrument noise is also shown for comparison (dotted curve).}  
\end{figure}

At first, the instrument noise of Taiji is generated from the analytic target model~\cite{Ruan:2018tsw, LUO2020102918}:
\begin{widetext}
\begin{eqnarray}
  P_{\rm dp} &=& \left(8 \times 10^{-12} \, {\rm m}\right)^2 \left( 1+ \left(\frac{ 2\, {\rm mHz}}{f}\right)^4 \right) {\rm Hz}^{-1} ,  \\
  P_{\rm acc} &=& \left(3 \times 10^{-15} \, {\rm m}\, {\rm s}^{-2}\right)^2 \left( 1+ \left(\frac{ 0.4\, {\rm mHz}}{f}\right)^2 \right) \left( 1+ \left(\frac{f}{ 8\, {\rm mHz}}\right)^4 \right)  \, {\rm Hz}^{-1}\, ,
\end{eqnarray}
\end{widetext}
where $P_{\rm dp}$ is the power spectral density(PSD) of the displacement noise and $P_{\rm acc}$ is the PSD of the acceleration noise.
In the source frame, the time domain waveform of a GB can be written as~\cite{Baghi:2022ucj, ldc}
\begin{eqnarray}
    h_{+}(t) &=&  \mathcal{A} \left(1+\cos^{2}\iota\right) \cos \left(\Phi(t)\right) , \\
    h_{\times}(t) &=& -2\mathcal{A} \cos \iota \sin \left(\Phi(t)\right) , \\
    \Phi(t) &=& \phi_0 + 2\pi f t +\pi \dot{f} t^2 ,
\end{eqnarray}
where $\mathcal{A}$ is the amplitude, $\iota$ is the inclination angle, $\Phi$ is the orbital phase of the binary, $\phi_0$ is the initial phase, $f$ and $\dot{f}$ is the frequency and the derivative of the frequency of GWs.
The technique of time delay interferometry (TDI) is proposed for space-based detection to suppress the laser frequency noise~\cite{1999ApJ...527..814A, Tinto:2020fcc, Wang:2017aqq}.
With the implementation of TDI, the signals from different channels are combined into the new TDI channels~\cite{Tinto:2020fcc}.
To generate the signals of 29.8 million GBs in the first generation TDI channels X, Y and Z, the rigid adiabatic approximation is used to calculate the Taiji response with a 4-year mission lifetime~\cite{Rubbo:2003ap, Cornish:2007if, Timpano:2005gm, Liu:2020mab}.
All the GW signals are added to the instrument noise to get the \textit{whole} signal whose PSD is estimated by using the BayesLine algorithm~\cite{Littenberg:2014oda}.
Based on the PSD, we calculate the signal-to-noise ratio (SNR) of each source and subtract the \textit{resolvable} sources whose SNR $> 7$ from the whole signal data.
Here we assume that the sources can be removed perfectly without residuals but in real data analysis this would not happen and the subtraction errors should be considered~\cite{Robson:2017ayy, Cornish:2003vj}.
After the subtraction, the PSD is re-estimated and updated. 
Then we repeat the procedure of subtracting resolvable sources and re-estimating the PSD. 
After 10 iterations, the number of the subtracted sources is less than $10$ and the PSD is almost unchanged.
The final PSD of the data containing the confusion noise and instrument noise is obtained, see Fig.~\ref{fig:pnpcX}.
In this figure, only the X channel is shown.
From Fig.~\ref{fig:pnpcX} it is clear that after removing the resolvable sources the residual confusion noise for Taiji is dominant over the instrument noise around 1~mHz, with a spectrum similar to the case for LISA~\cite{Cornish:2017vip, Karnesis:2021tsh}. 

\begin{figure}
  \includegraphics[width=0.9\linewidth]{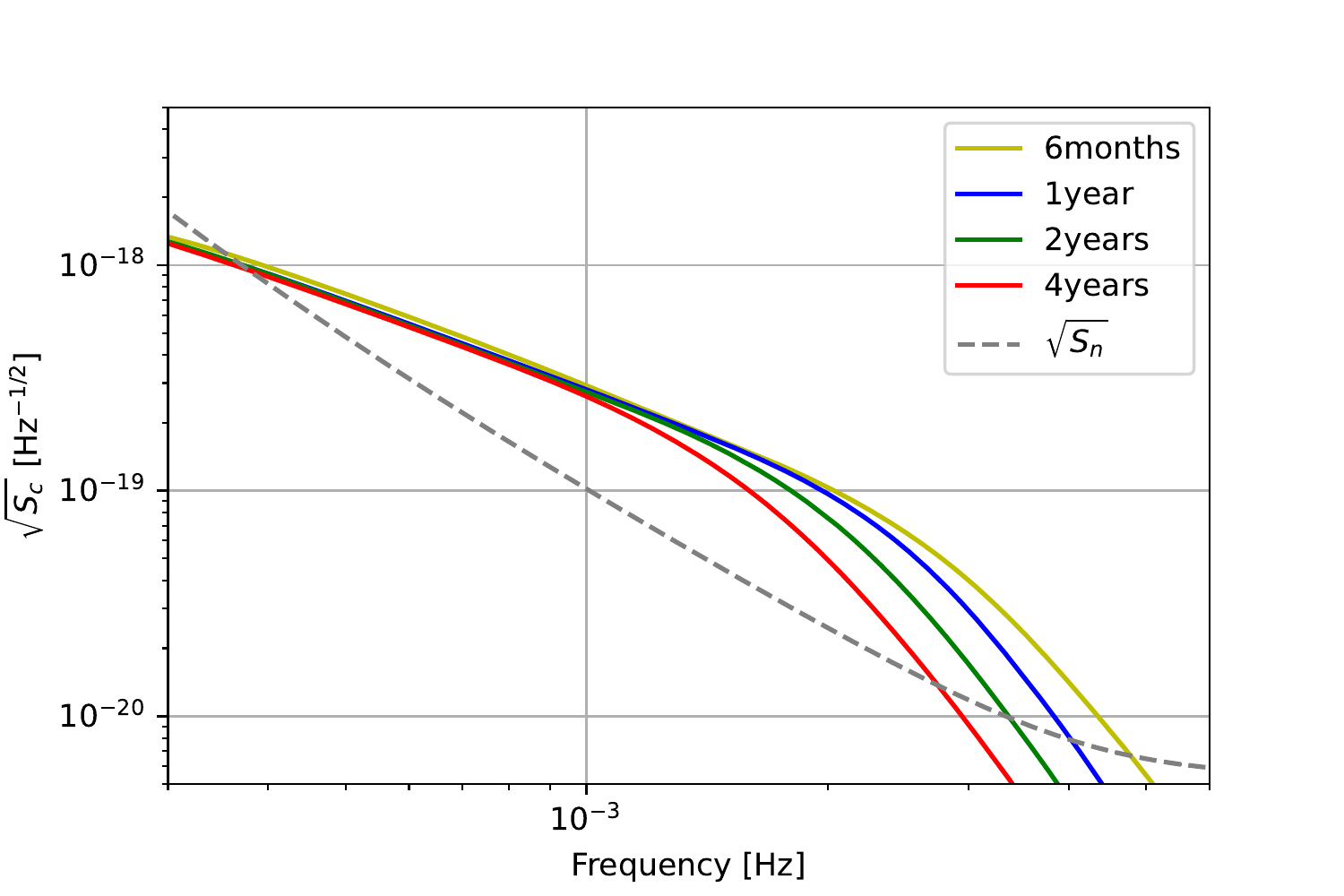}
  \caption{\label{fig:scsTaiji} The confusion noise for different observation time. The dashed grey curve represents the sensitivity curve of Taiji.}
\end{figure}

We also simulate the first-generation TDI signals for different mission durations.
To get the confusion noise, we convert the spectrum of XYZ channels into the effective noise PSD in the sensitivity curve~\cite{Robson:2018ifk, Cornish:2007if}.
Fig.~\ref{fig:scsTaiji} shows the confusion noise for different observation time $T_{obs}=$ 6~months, 1~year, 2~years and 4~years. 
The number of resolvable sources increases with observation time and the level of confusion noise goes down, especially for frequencies above 1~mHz.
We use a polynomial function to fit the confusion noise $S_c(f)$ in the logarithmic scale as
\begin{equation}
  S_c(f)=\exp\left(\sum_{i=0}^{5} a_i \left(\log\left(\frac{f}{\rm mHz}\right)\right)^i \right) \, {\rm Hz}^{-1}\, .
  \label{eq:sc}
\end{equation}
The fitting only works for $0.1 \, {\rm mHz} < f < 10 \, {\rm mHz}$ and the parameters $a_i$ for different observation time are shown in Table~\ref{tab:ai}.
\begin{table*}
  \centering
  \caption{\label{tab:ai} Fitting parameters of the confusion noise $S_c(f)$ in Eq.~\eqref{eq:sc} for different observation time $T_{obs}$.}
  \begin{tabular}{c c c c c c c}
  \hline
  $T_{obs}$ & $a_0$ & $a_1$ & $a_2$ & $a_3$ & $a_4$ & $a_5$ \\
      \hline
      6 months & -85.3498 & -2.64899 & -0.0699707 & -0.478447 & -0.334821 & 0.0658353 \\
      1 year & -85.4336 & -2.46276 & -0.183175 & -0.884147 & -0.427176 & 0.128666 \\
      2 years & -85.3919 & -2.69735 & -0.749294 & -1.15302 & -0.302761 & 0.175521 \\
      4 years & -85.5448 & -3.23671 & -1.64187 & -1.14711 & 0.0325887 & 0.187854 \\
      \hline
  \end{tabular}
\end{table*}

\begin{figure}
  \includegraphics[width=0.9\linewidth]{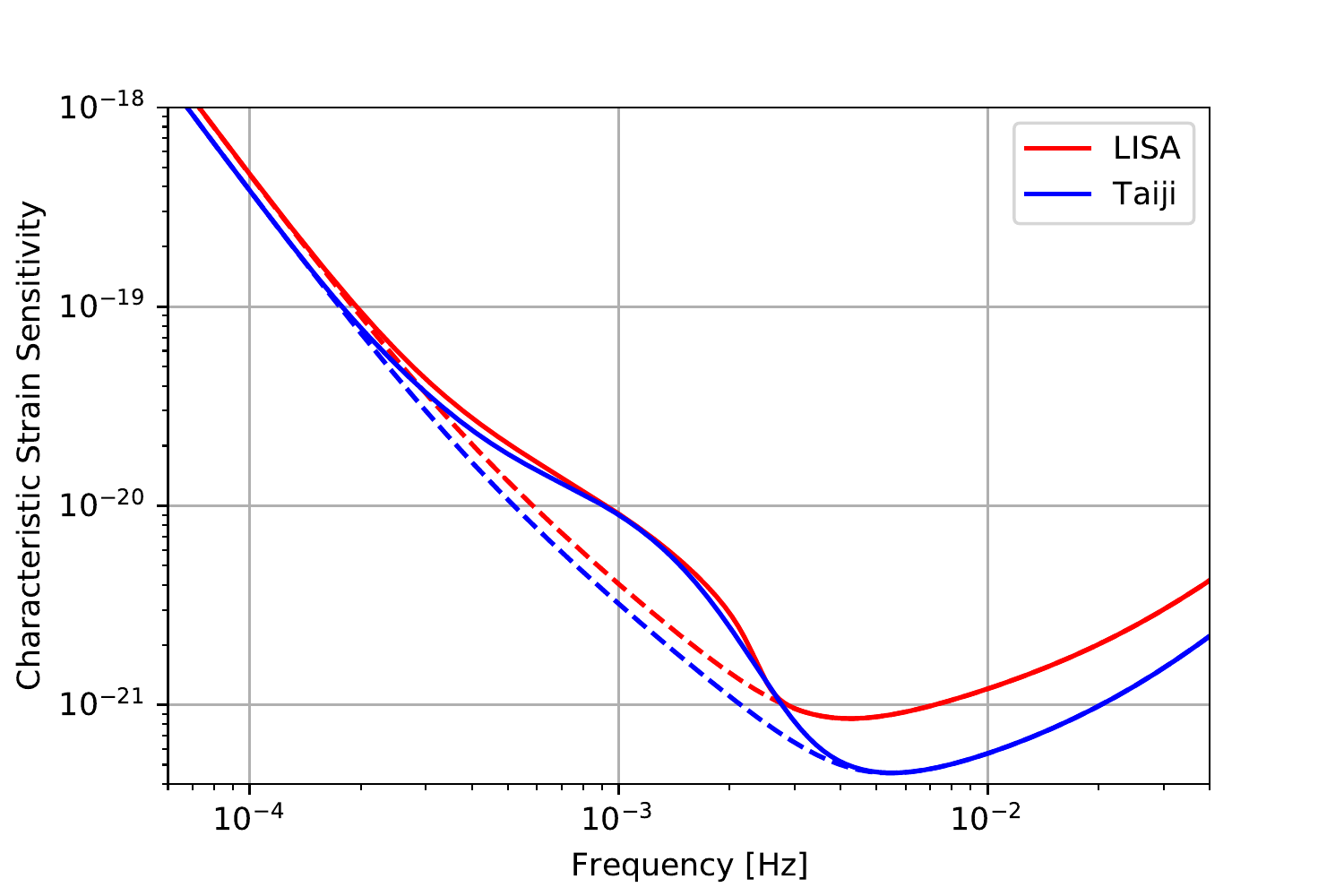}
  \caption{\label{fig:sensiLT} The sensitivity curves of Taiji and LISA. The dashed curves represent the design sensitivities. The confusion noises for the 4-year observation are included in the solid curves to show the full sensitivity curves. Here we plot the dimensionless characteristic strain sensitivity $\sqrt{f S_n}$.}
\end{figure}

In Fig.~\ref{fig:scsTaiji}, the design sensitivity curve $S_n$ of Taiji is also plotted for comparison. For space-based GW detectors, such as LISA and Taiji, the sensitivity curve $S_n$ for the two Michelson-style data channels can be written as~\cite{Robson:2018ifk}
\begin{eqnarray}
  S_{n}(f) &=& \frac{10}{3 L^2} \left( P_{\rm dp}  + 2(1+\cos^2(f/f_*)) \frac{P_{\rm acc}}{(2\pi f)^4} \right) \nonumber \\
  && \times \left(1 + 0.6\left(\frac{f}{f_*}\right)^2 \right) \, , 
  \label{eq:Sn}
\end{eqnarray}
where $f_*=1/(2 \pi L)$ and $L$ is the armlength. For Taiji $L=3\times 10^9$ m. To include the confusion noise one needs to add the $S_c$ to the sensitivity curve $S_n$. Figure~\ref{fig:sensiLT} shows the full sensitivity curves of Taiji and LISA~\cite{Robson:2018ifk}.
The noise model of LISA can be found in the LISA Science Requirements Document~\cite{SciRD} or Ref.~\cite{Robson:2018ifk}. For LISA, the armlength is $2.5\times 10^9$ m and the displacement noise 
\begin{equation}
  P_{\rm dp L} = \left(15 \times 10^{-12} \, {\rm m}\right)^2 \left( 1+ \left(\frac{ 2\, {\rm mHz}}{f}\right)^4 \right) {\rm Hz}^{-1} \, .
\end{equation}
The acceleration noise is the same as that of Taiji.
To generate the full sensitivity curve of LISA, we use the empirical fitting model of the confusion noise for LISA from Ref.~\cite{Karnesis:2021tsh}.
In Fig.~\ref{fig:sensiLT} we can see that the confusion noise is slightly weaker for Taiji than for LISA at the frequency of $\leq 0.8$ mHz and around 2~mHz. 
This is due to the fact that Taiji's arm length is longer than LISA's and the instrument noise level is slightly lower than LISA's. 
But at the frequency of $\sim 1$~mHz, the confusion noise is nearly identical for both.
Because it is much stronger than the instrument noise at 1~mHz, the effect of the different configurations is negligible.

\section{\label{sec:resolvable}Resolvable sources}

In our analysis, the sources whose SNR $> 7$ are marked as \textit{resolvable} as in Ref.~\cite{Cornish:2017vip, Karnesis:2021tsh}. 
Table~\ref{tab:num} lists the number of resolvable sources for different observation time.
\begin{table}
  \centering
  \caption{\label{tab:num}The number of resolvable GBs with Taiji and LISA for different observation time $T_{obs}$.}
  \begin{tabular}{c c c}
    \hline
      $T_{obs}$ & Taiji  & LISA \\
      \hline
      6 months & 7083  & 4697 \\
      1 year & 11439  & 8830 \\
      2 years & 18500  & 14939 \\
      4 years & 29633  & 24780 \\
    \hline
  \end{tabular}
\end{table}
With increasing observation time, more sources become resolvable.
As pointed out in Ref.~\cite{Karnesis:2021tsh}, using different smoothing methods to estimate the PSD, one will get slightly different results.
In Ref.~\cite{Karnesis:2021tsh}, two methods were used to smooth the PSD: running median and running mean. 
Here we use a different one: the BayesLine algorithm, which is the same as in Ref.~\cite{Cornish:2017vip}.
To compare with our results for Taiji, we perform the same analysis for LISA following the procedure described in Sec.~\ref{sec:sctaiji} and obtain the resolvable sources, as shown in Table~\ref{tab:num}. 
The number of resolvable sources for LISA is consistent with the results of Ref.~\cite{Karnesis:2021tsh}.
It is larger than the number from the running median method and smaller than the result from the running mean method (see Table I of Ref.~\cite{Karnesis:2021tsh}).
As shown in Table~\ref{tab:num}, Taiji allows one to subtract $\sim 20 \%$ more sources than LISA.
This is because the instrument sensitivity of Taiji is better than LISA's.
A recent work~\cite{Zhang:2022wcp} has shown that the network of two space-based detectors can resolve $\sim 75 \%$ more confirmed sources than a single one.

\begin{figure}
    \includegraphics[width=0.9\linewidth]{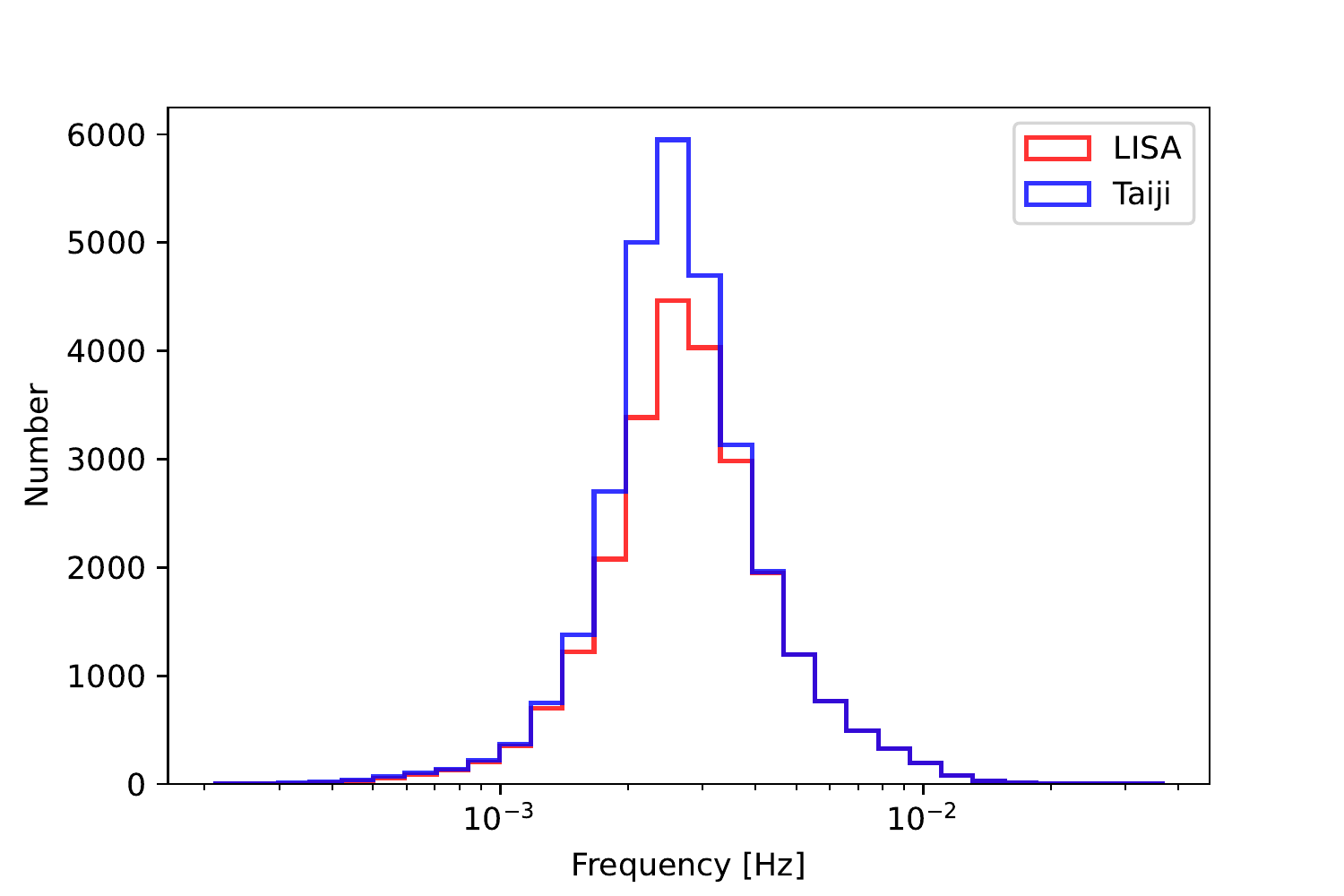}
    \includegraphics[width=0.9\linewidth]{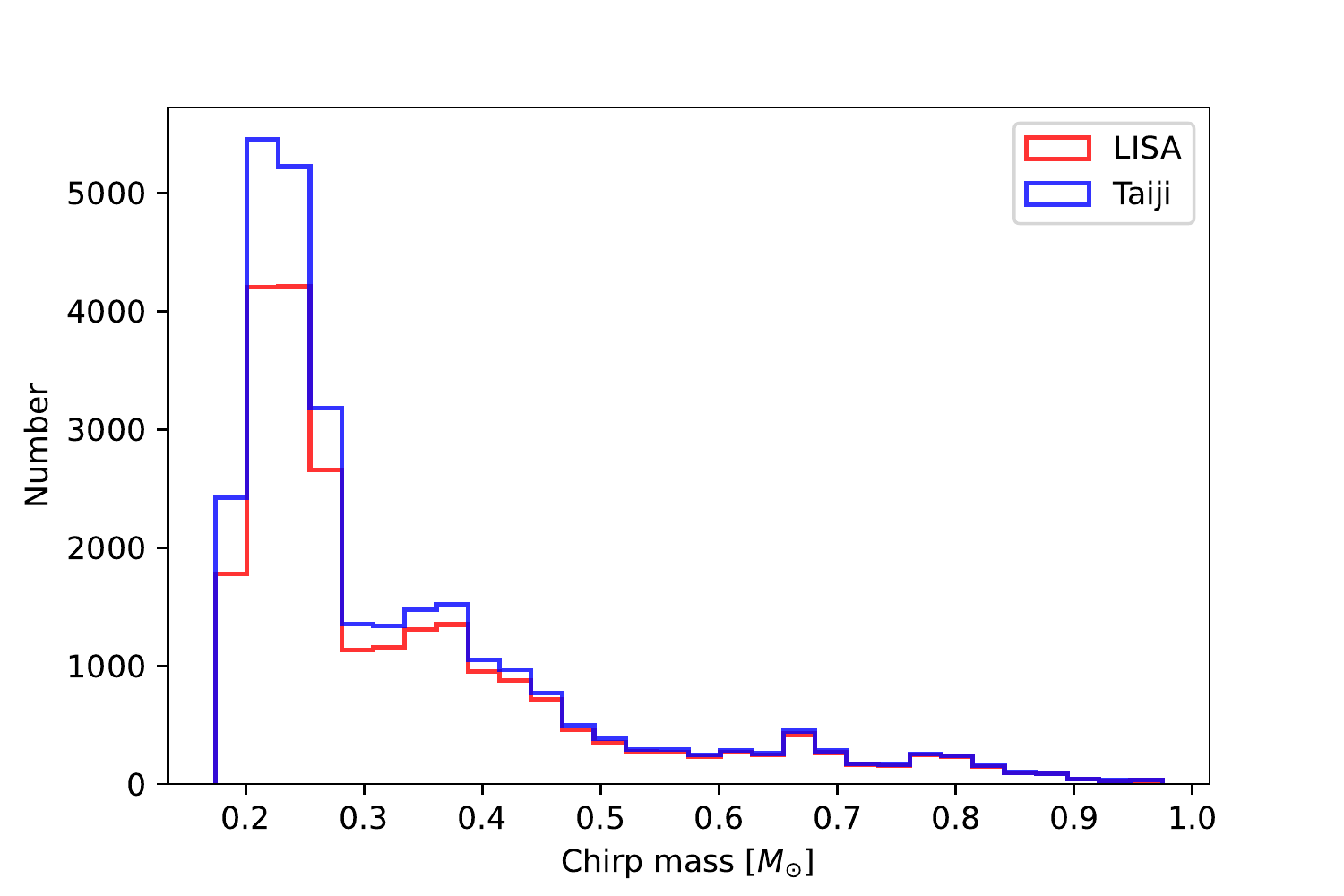}
    \includegraphics[width=0.9\linewidth]{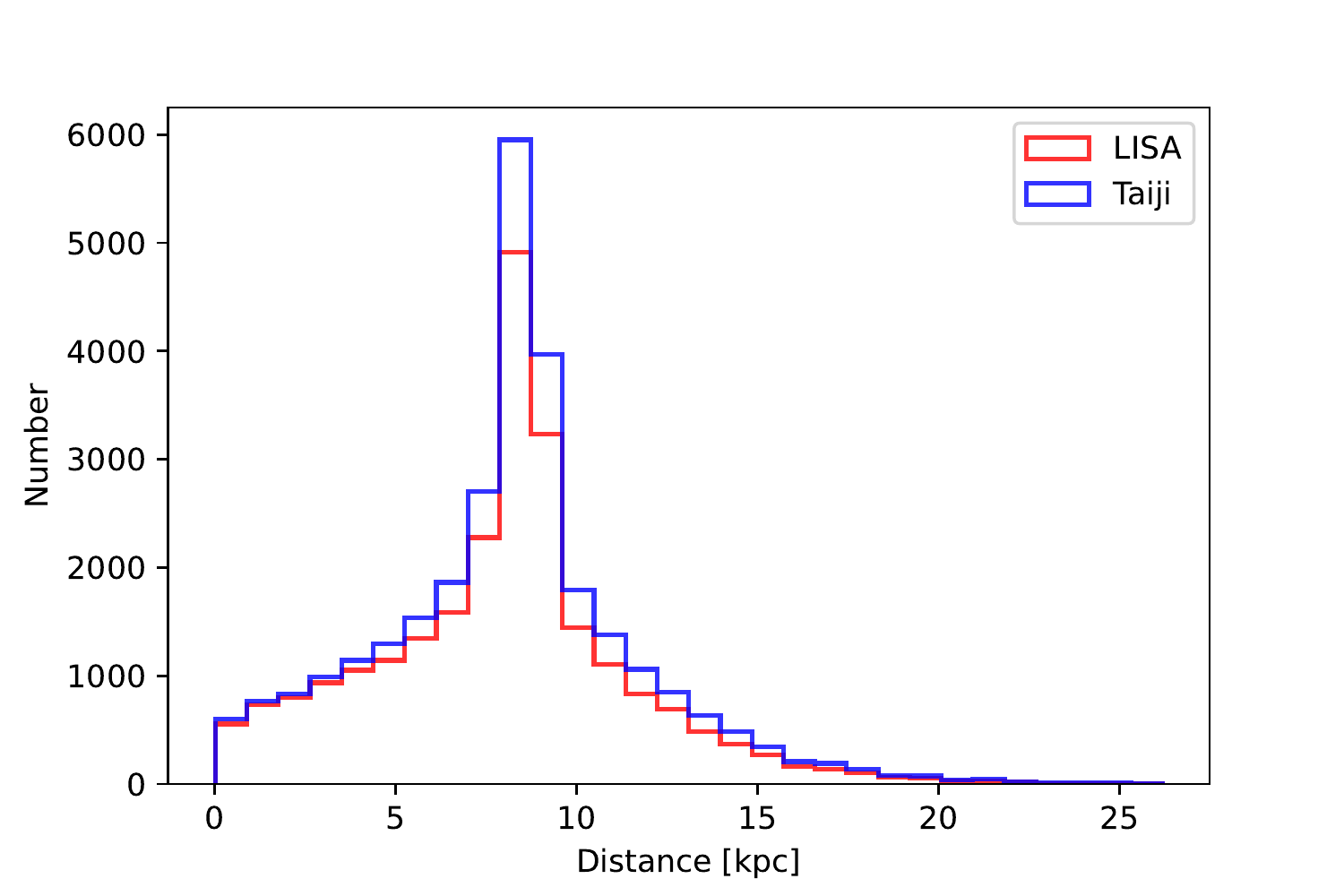}
    \caption{\label{fig:resolvablehist} The distributions of the GW frequencies, chirp masses, and distances of the detected GBs resolvable with Taiji and LISA for a 4-year observation.}
\end{figure}

In Fig.~\ref{fig:resolvablehist}, we show the distributions of the GW frequencies, chirp masses, and distances of the detected GBs resolvable with Taiji and LISA.
Taiji can subtract more sources than LISA around 2~mHz, at which Taiji has slightly better sensitivity.
The middle panel of Fig.~\ref{fig:resolvablehist} shows that more sources with the chirp masses ranging from  $0.2 M_\odot$ to $0.4 M_\odot$ can be resolvable with Taiji.
The number of GBs is relatively large near the center of the Milky Way ($\sim 8$ kpc), where more sources are resolvable with Taiji (see the bottom panel of Fig.~\ref{fig:resolvablehist}).

Moreover, we find that all the sources resolvable with LISA are also resolvable with Taiji. 
\begin{figure*}
    \includegraphics[width=0.8\linewidth]{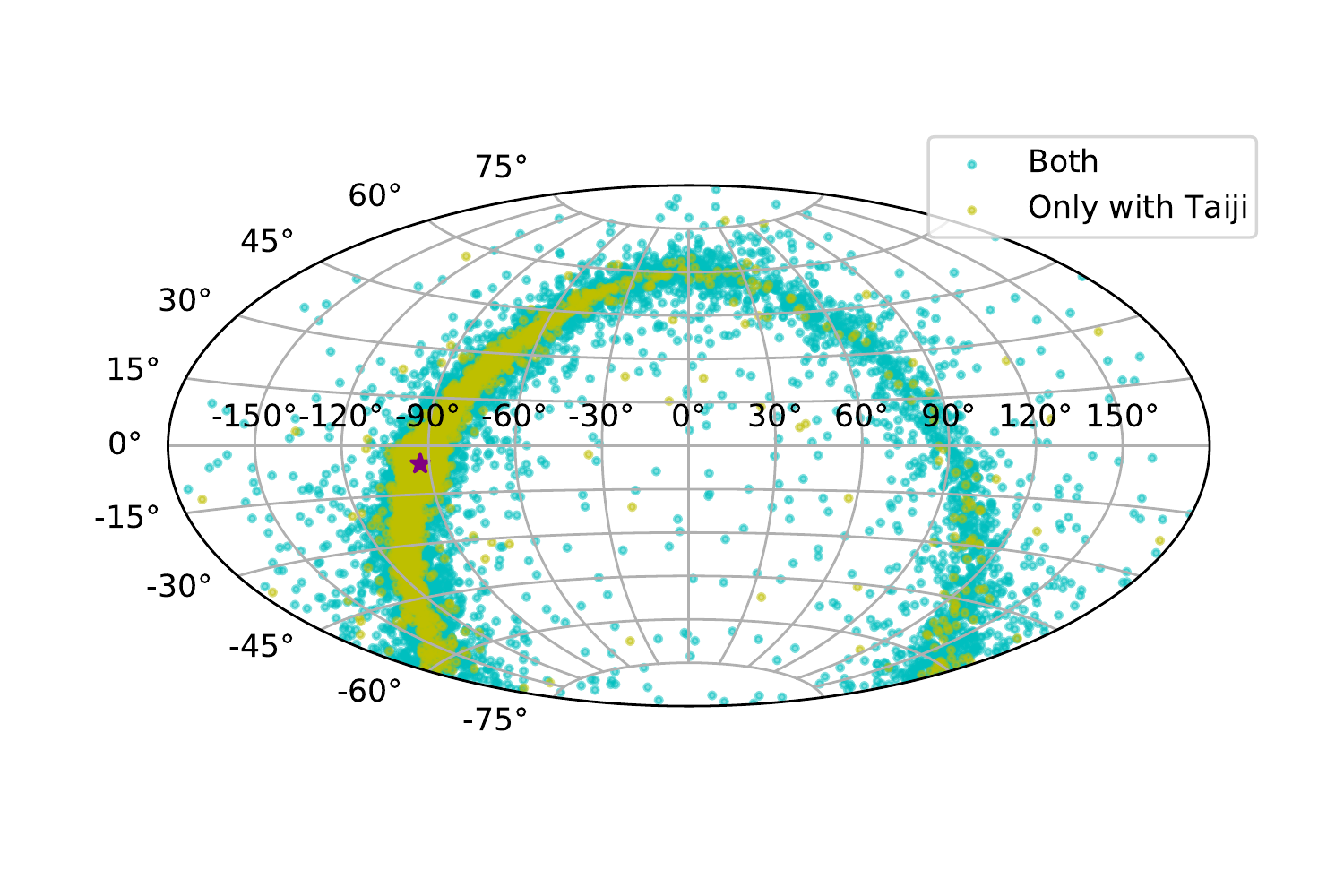}
    \caption{\label{fig:resolvableskymap} Sky positions of the resolvable sources in the ecliptic coordinate system for a 4-years observation. The cyan points correspond to the sources resolvable with LISA and Taiji and the yellow points indicate the sources only resolvable with Taiji. The Galactic Center is marked by the purple star.}
\end{figure*}
Fig.~\ref{fig:resolvableskymap} shows the sky positions of the resolvable sources for a 4-year observation.
The cyan points correspond to the 24780 sources resolvable with LISA and Taiji and the yellow points correspond to the 4853 sources only resolvable with Taiji.
Since Taiji and LISA have similar constellation designs, the distribution of the yellow points and the cyan points in Fig.~\ref{fig:resolvableskymap} are very similar.

The performance of the parameter estimation for the resolvable sources with Taiji is an interesting topic for future work.

\section{\label{sec:summary}Summary}

We use the catalog of 29.8 million GBs provided by LDC to simulate the foreground signals for Taiji. 
The sources with SNR larger than 7 are treated as \textit{resolvable} and can be subtracted from the data.
With Taiji for a 4-year lifetime, more than 29 thousand sources are resolvable and the residual signal produces an effective confusion noise.
For different observation time $T_{obs}=$ 6~months, 1~year, 2~years and 4~years, we fit the confusion noise $S_c(f)$ by using polynomial functions on a log-log scale.
With longer observation time, the confusion noise becomes significantly lower at the frequency above 1~mHz, as in the case of LISA~\cite{Cornish:2017vip}.
To get the full sensitivity of Taiji, the confusion noise $S_c$ should be added to the instrument noise.
The full sensitivity curve is slightly lower for Taiji than for LISA at the frequency of $\leq 0.8$ mHz and around 2~mHz.

The number of resolvable sources increases with the duration of observation.
Compared to LISA, Taiji can subtract $\sim 20 \%$ more sources.
Their distribution in our Milky Way is consistent with that of the resolvable sources with LISA.
And at frequencies around 2~mHz or with the chirp masses ranging from  $0.2 M_\odot$ to $0.4 M_\odot$, more sources become resolvable with Taiji.
Taiji can subtract more sources near the Galactic Center.

Here we assume perfect subtraction of the resolvable signals which means that the parameters of the sources can be perfectly estimated so that the true waveform can be removed from the data stream.
For real data analysis, imperfect subtraction will introduce residuals into the data~\cite{Robson:2017ayy, Cornish:2003vj}.
In this case, the \textit{global fit} is required which models all the sources together for parameter estimation~\cite{Cornish:2005qw, Littenberg:2020bxy}.

In addition, we regard the confusion noise to be stationary across different observation periods.
In reality, as the space-based detector rotates around the sun, the response function concerning the sources in the Milky Way varies over time.
As a result, the galactic foreground varies throughout of the year and yields a cyclostationary noise~\cite{Seto:2004ji, Digman:2022jmp}.
We shall add this effect in our future work for Taiji.

The catalog of GBs we used is based on binary population synthesis provided by LDC~\cite{Baghi:2022ucj, Korol:2020lpq}.
Recent studies based on the observationally driven population indicate that the shape of the confusion noise will be different~\cite{Korol:2021pun}.
The updated population can be included in future work.

By now, since all the planned space-based detectors have not yet been launched, the data analysis methods are developed by using mock data. 
The Taiji confusion noise we obtain here will assist researchers in investigating the capability of more realistic data analysis methods for Taiji mission.

\begin{acknowledgments}
CL is supported by the National Natural Science Foundation of China under Grant No. 12147132. 
ZKG is supported by the National Natural Science Foundation of China Grants No.12235019 and No.12075297.

We thank Jing Liu, He Wang and Yuan-Hao Zhang for useful discussions and comments.
The authors would like to acknowledge the work of the LDC group.
For this study, both the LDC code and datasets were used~\cite{ldc}.
Part of the code from LDASoft was used~\cite{Littenberg:2020bxy, ldasoft}.
The authors also thank Neil Cornish and Travis Robson for sharing their code to do the analysis for LISA~\cite{Cornish:2017vip}.
This work made use of NumPy~\cite{harris2020array}, SciPy ~\cite{2020SciPy-NMeth}, Astropy~\cite{2022ApJ...935..167A} and Matplotlib~\cite{Hunter:2007}.
\end{acknowledgments}

\bibliography{mybib.bib}

\end{document}